\documentstyle[12pt]{article}
\newcommand{\be}{\begin{equation}}
\newcommand{\ee}{\end{equation}}
\newcommand{\bea}{\begin{eqnarray}}
\newcommand{\eea}{\end{eqnarray}}
\begin{document} 

\baselineskip=18pt
\begin{center}
{\Large{\bf 
Elementary Excitations of a Relativistic Scalar \\
Plasma System }}
\end{center}
\vspace{0.3cm}
\baselineskip=15pt
\begin{center}
E.R. Takano Natti$^{\dag}$, Chi-Yong Lin$^{\dag}$ and A.F.R. de Toledo Piza
\end{center}
\begin{center}
{\it Instituto de F\'{\i}sica, Universidade de S\~ao Paulo \\
     Caixa Postal 66318, 05315-970, S\~ao Paulo\\
     S\~ao Paulo \ \ SP \ \ \ Brazil} 
\end{center}
\begin{center}
P.L. Natti 
\end{center}
\begin{center}
{\it Departamento de Matem\'atica, Universidade Estadual de Londrina,\\
Caixa Postal 6001, Cep 86051-970, Londrina \ \  Paran\'a \ \ \  Brazil}
\end{center}

\baselineskip=15pt
\vspace{0.5cm}
\begin{center}
{\bf ABSTRACT}
\end{center}

\indent
We investigate the physics of elementary excitations for the so called
relativistic scalar plasma system. 
Following the standard many-body procedure we have obtained the RPA
equations for this model by linearizing the TDHFB equations of motion
around equilibrium and shown that these oscillation modes give
one-meson and two-fermion state of the theory. 
The resulting equations have a closed solution, from which one can
examinate the spectrum of excitation modes. 
In partticular, our results indicate existence of bound state for
certain region of phase diagram.

\vskip 1.0cm
\vskip 1.0cm

\begin{center}
\end{center}

\vfill

\noindent\makebox[66mm]{\hrulefill}

\footnotesize 

\noindent
$^{\dag}$
Supported by Conselho Nacional de Desenvolvimento Cient\'{\i}fico e
Tecnol\'ogico (CNPq), Brazil.

\normalsize
\newpage

\normalsize
\newpage
\baselineskip=21pt
\renewcommand{\theequation}{1.\arabic{equation}}
\setcounter{equation}{0}
\begin{center}
{\bf I. INTRODUCTION}
\end{center}

\indent
In a previous work \cite{TNa98} (hereafter referred to as I) we have
presented a framework to investigate the initial-value problem in the
context of interacting fermion-scalar field theories. 
The method allows one to obtain a set of self-consistent equations for
the expectation values of linear and bilinear forms of field
operators.
These dynamical equations acquire the structure of kinetic type,
where the lowest order of the approximation corresponds to the usual
gaussian mean-field approximation (collisionless).   
As application, we have implemented a zero-order calculation within
the simplest context of relativistic scalar plasma system.
We have shown that the usual prescription of renormalization can also
apply to these nonoperturbative calculation. 
In particular, we have obtained a finite expression for the energy
density and the numerical results suggested that the system
presents always a single stable minimum .

\indent
In continuation of I we will report in this paper a particular
application of the renormalized nonlinear obtained the previous
publication. 
We follow here a recent work by Kerman and Lin \cite{KL95,KL98} in
order to investigate the near equilibrium dynamics around the
stationary solution. 
We shall show that one-meson and two-(quasi)fermion physics can be
studied from the linear approximation of the mean-field equations. 
In particular, one can solve these equations in a closed form and find
scattering amplitude as well as the conditions for the two-fermion
bound state.

\indent
For completeness and notational purpose, we repeate here a few key
equations of I.
A summary of derivation for these equations is shown in Appendix A. 
For the scalar plasma model, the dynamics are governed by  
the hamiltonian 

\begin{eqnarray}
\label{1}
H&=&\int_{\bf x} {\cal H}\;,\nonumber\\ \\ {\cal H}&=&-\bar\psi
(i\vec\gamma.{\vec\partial}-m)\psi-
g\bar\psi\phi\psi+\frac{1}{8\pi}
\left[\frac{(4\pi)^2}{1+Z} \Pi^2+(1+Z)|\partial\phi|^2+
\mu^2\phi^2\right]+{\cal H}_{c}\;\;, \nonumber
\end{eqnarray}
\vskip 0.5cm
\noindent 
[We use the notation: $\int_{\bf x}=\int d^{3}x$]
where $\psi$ is a spin-$\frac{1}{2}$ field while $\phi$ is a scalar
field. The parameters $m$ and $\mu$ are, respectively, the mass of
fermion and scalar particles and $g$ is the coupling constant. The
last term of this expression is the counterterms necessary to remove
the infinities occurring later \cite{TNa98,AH88}, 

\begin{equation}
\label {27}
4\pi {\cal H}_{\mbox {\tiny C}} = \frac{A}{1!}\phi + 
\frac{\delta \mu^{2}}
{2!}\phi^{2} + \frac{C}{3!}\phi^{3} + \frac{D}{4!}\phi^{4} \;\;,
\end{equation}
\vskip 0.5cm
\noindent
where the coeffcients $A$, $\delta \mu^{2}$, $C$, $D$, and $Z$ are
infinity constants to be definided later. 

\indent
In order to study the dynamics of the system, we focus on the selected
set of observables, which are related to the expectation values of
linear and bilinear forms of field operators, referred to as
gaussian variables.
The time evolution of these quantities obeys the Heisenberg equation of
motion,

\begin{equation}
i\langle \dot{\cal O} \rangle 
=Tr_{\mbox{\tiny BF}}[{\cal O},H] {\cal F}
\end{equation}

\noindent 
where ${\cal O}$ could be $\phi(x)$, $\phi(x)\phi(x)$,
$\bar\psi(x)\psi(x)$ and etc., and $F$ is the state of the system in
the Heisenberg picture. 
%
As approximation we replace the full density $\cal F$ by a truncated
ansatz ${\cal F}_{0}(t)$.
By construction, ${\cal F}_{0}$ reproduces the corresponding ${\cal F}$
averages for linear or bilinear field operators [see Eqs.(41) of
\cite{TNa98}]. 
In particular, we have used a formulation appropriate for the
many-body problem, so that 
${\cal F}_{0}$ can be written in the momentum basis as \cite{LTP92,NTP96}

\begin{eqnarray}
{\cal F}_{0} 
&=&{\cal F}^{\mbox{\tiny B}}_{0} {\cal F}^{\mbox{\tiny F}}_{0}\nonumber\\
\label{32a}
{\cal F}^{\mbox{\tiny B}}_{0}
&=&\prod_{\bf p}\frac {1} {1+\nu_{\bf
p}}\left(\frac {\nu_{\bf p}}{1+\nu_{\bf p}}\right) ^{\beta_{\bf
p}^{\dag}\beta_{\bf p}} \\
\label{32b}
{\cal F}^{\mbox{\tiny F}}_{0}
&=&\prod_{{\bf k},s,\lambda} [\nu_{{\bf
k},s}^{(\lambda)}{\alpha^{(\lambda)}_{{\bf k},s}}^{\dag}
\alpha^{(\lambda)}_{{\bf k},s}+(1-\nu_{{\bf k},s}^{(\lambda)})
\alpha_{{\bf k},s}^{(\lambda)}{\alpha^{(\lambda)}_{{\bf
k},s}}^{\dag}]\;\;,
\end{eqnarray}
\vskip 0.3cm
\noindent 
where $\alpha$ ($\alpha^{\dag}$) and $\beta$ ($\beta^{\dag}$) stand for
Bogoliubov quasi-particle annhihilation (creation) operators for fermion
and boson respectively. In doing so, the gaussian variables are now
represented by the Bogoliubov parameters [see (\ref{4})-(\ref{5})] and its
equations of motion, 
can be obtained directly using
(\ref{1})-(\ref{32b}). The resulting expressions are:

\begin{eqnarray}
\label{41}
\dot\varphi_{{\rm k}}&=&g\langle\phi\rangle\frac{|{\bf k}|}{k_{0}}\sin
\gamma_{{\rm k}}\\\nonumber\\ 
\label{42}
\dot\gamma_{{\rm k}}&=&2k_{0}-2gm
\langle\phi\rangle\frac{1}{k_{0}}\left(1-\frac{|{\bf k}|}{m}\cot
  2\varphi_{{\rm k}}\cos\gamma_{{\rm k}}\right)\\\nonumber\\ 
\label{43} 
\langle\dot\phi\rangle&=&\frac{4\pi}{(1+Z)}\langle\Pi\rangle\\\nonumber\\
\label{44}
\langle\dot\Pi\rangle&=&-\left(\frac{A}{4\pi}+\frac{C}{2}G(\mu
)\right)-\left(\frac{\mu^2}{4\pi}+\frac{\delta\mu^2}{4\pi}+\frac{D}{2}G(\mu)
\right)\langle\phi\rangle\nonumber\\\nonumber\\
&-&\frac{C}{8\pi}\langle\phi\rangle^2-\frac{D}{24\pi}\langle\phi\rangle^3
-\frac{2g}{(2\pi)^3}\left[I_{1}(m)+I_{2}(m)\right]
\\\nonumber
\end{eqnarray}
\vskip 0.3cm
\noindent
where 

\begin{eqnarray}
\label{45}
G(\mu)&=&\frac{1}{(2\pi)^3}\int d^3{\rm p}\frac{1}{2({\bf p}^2+\mu^2)^{1/2}}\\\nonumber\\
\label{46}
I_{1}(m)&=&\int d^3{\rm k}\frac{m}{({\bf k}^2+m^2)^{1/2}}\cos 2\varphi_{\rm k}\\\nonumber\\
\label{47}
I_{2}(m)&=&\int d^3{\rm k}\frac{|{\bf k}|}{({\bf k}^2+m^2)^{1/2}}\sin 2\varphi_{\rm k}\cos\gamma_{\rm k}\;\;.\\\nonumber
\end{eqnarray}
\vskip 0.3cm
\noindent
The mean-field energy density, on the other hand, reads as

\begin{eqnarray}
\label{123a}
&&\frac{\langle H\rangle}{V} = \frac{1}{V}Tr H{\cal F}_{0}
\nonumber\\\nonumber\\
&&=-2\int_{\rm k}
k_{0}\cos 2\varphi_{{\rm k}}
+2g\langle\phi\rangle 
\left[I_{1}(m)+I_{2}(m)\right]
\nonumber\\\nonumber\\
&&+\frac{1}{4\pi}\left(\frac{\mu^2}{2}\langle\phi\rangle^2+
\frac{\langle\Pi\rangle^2}{2}\right)
+\left(\frac{A}{4\pi}+
\frac{C}{2}G(\mu)\right)
\langle\phi\rangle
\nonumber\\\nonumber\\
&&+\left(\frac{\mu^2}{8\pi}+\frac{\delta\mu^2}{8\pi}+
\frac{D}{4}G(\mu)\right)
\langle\phi\rangle^2
+\frac{C}{24\pi}\langle\phi\rangle^3+
\frac{D}{96\pi}\langle\phi\rangle^4\;\;
\end{eqnarray}
\vskip 0.3cm

\indent The equations above describe the real-time evolution for the
relativistic scalar plasma system in the gaussian mean-field approximation.  
They are nonlinear time-dependent field equations. Therefore,
a closed solution is not easily constructed. Here, we will 
consider the equations in the equilibrium situation and the small
oscillation regime. In these cases, a closed solution can be found
allowing us to examine diverse properties of the theory.
The structure of this paper is as follows.  In Sec. II we shall derive
the RPA equation for this model by considering near equilibrium
dynamics about the equilibrium solutions obtained in I. Section III
shows an analytical solutions for these RPA equations using a well
know procedure of the scattering theory.
In Section IV we discuss the question of renormalization within the
context of scattering amplitudes and examinate the possibility of
existence of bound state soultions.

\renewcommand{\theequation}{2.\arabic{equation}}
\setcounter{equation}{0}
\begin{center}
{\bf II. NEAR EQUILIBRIUM DYNAMICS }
\end{center}

\indent
As a function of the Bogoliubov parameters, the energy density
is a function of these variables.
A minimum of (\ref{123a}) corresponds to the ground state of the system
and small amplitude motion about the minima is described  by the
linearized approximation of the gaussian equations
motion (\ref{41})-(\ref{44}), yielding a set of harmonic oscillators
\cite{KK76}. 
The normal modes and the eigenvectors are the RPA solutions. 
In the field theoretical context these can be seen as the energy
and the wavefunctions of quantum particles. 
In this section we derive the RPA equations of the model and next
section will discussion solutions for this problem.

\indent
Let us consider first the static problem (see section VI-b of I).
Recalling (\ref{41})-(\ref{44}), the stationary conditions require
$
\dot\gamma_{\rm k}=\dot\varphi_{\rm k}= \dot\kappa_{\rm p}
=\dot\eta_{\rm p}=\langle\dot\phi\rangle= \langle\dot\Pi\rangle=0.
$
In Section IV-b of I we have discussed in detail the renormalization
conditions and the solutions and for this set of equations. 
In particular, the following choice for the coefficients of ${\cal
  H}_{c}$ satifies the self-consistency condition \cite{TNa98,AH88}

\begin{eqnarray}
\label{94}
D&=&\pm48\pi g^4 L(m)\;,\\
\label{95}
\delta\mu^2&=&\mp24\pi^2 g^4L(m)G(\mu)\mp16\pi g^2G(0)\pm24\pi m^2g^2 L(m)\;,\\
\label{96}
C&=&\mp48\pi mg^3L(m)\;,\\
\label{97}
A&=&\pm24\pi mg^3L(m)G(\mu)\pm16\pi mgG(m)\;,
\end{eqnarray}

\noindent
with

\begin{equation}
\label{98}
L(m) \equiv \int_{\bf k} \frac{1}{2{\bf k}^2 ({\bf k}^2+m^2)^{1/2}}\;\;.
\end{equation}
\vskip 0.3cm
\noindent 
In doing so, the finite static equations of the system can be regrouped
as follows: 

\begin{eqnarray}
\label{35}
&&\sin\gamma_{\rm k}|_{\mbox{\tiny eq}}=0\\\nonumber\\
\label{36}
&&\cot 2\varphi_{\rm k}|_{\mbox{\tiny eq}}=-\frac{({\bf k}^{2}+{\bar m}m)}
{|{\bf k}|(m-{\bar m})}\\\nonumber\\
\label{39}
&&\langle\Pi\rangle|_{\mbox{\tiny eq}}=0\\\nonumber\\
\label{40}
&&\frac{\pi}{2}{\mu}^2\langle\phi\rangle|_{\mbox{\tiny eq}}
-g{\bar m}^3\left[\ln\left(\frac{\bar m}{m}\right)+\frac{1}{2}\right]=0\;\;.
\\\nonumber
\end{eqnarray}
\vspace{0.3cm}
\noindent
where $\bar m$ is the effective fermion mass,

\begin{equation}
\label{86}
\bar m\equiv m-g\langle\phi\rangle|_{\mbox{\tiny eq}}\;\;.
\end{equation}

\noindent
The equation (\ref{35})-(\ref{40}) can be solved numerically for any
given values of $\mu$ and $g$ in unit of $m$. 


\indent
Investigation of the near equilibrium motion proceeds by considering
flutuactions around the stationary point

\begin{eqnarray}
\label{48}
\varphi_{\rm k}&=&\varphi_{\rm k}^{(0)}+\delta\varphi_{\rm
  k}\nonumber\\\nonumber\\ 
\gamma_{\rm k}&=&\gamma_{\rm k}^{(0)}+\delta\gamma_{\rm k}\nonumber\\\\
\langle\phi\rangle&=&\langle\phi\rangle^{(0)}+\delta\langle\phi\rangle\nonumber\\\nonumber\\
\langle\Pi\rangle&=&\langle\Pi\rangle^{(0)}+\delta\langle\Pi\rangle\;\;,\nonumber\\\nonumber
\end{eqnarray}

\noindent 
where $\varphi_{\rm k}^{(0)},\gamma_{\rm
    k}^{(0)},\langle\phi\rangle^{(0)}$ and $\langle\Pi\rangle^{(0)}$
    satisfy (\ref{35})-(\ref{40}) and
the quantities $\delta\varphi_{\rm k}$, $\delta\gamma_{\rm k}$,
$\delta\langle\phi\rangle|_{\mbox{\tiny eq}}$ and
$\delta\langle\Pi\rangle|_{\mbox{\tiny eq}}$ will be assumed
small in our approximation. 
Next step is to expand (\ref{41})-(\ref{44}) up to first order in
these fluctuations  to yield

\begin{eqnarray}
\label{54}
\delta\dot\varphi_{\rm k}&=&g\langle\phi\rangle^{(0)}\frac{|{\bf k}|}{k_{0}}\delta\gamma_{\rm k}\\\nonumber\\
\label{55}
g\langle\phi\rangle^{(0)}|{\bf k}|\delta\dot\gamma_{\rm k}&=&-4k_{0}({\bf k}^2+M^2)\delta\varphi_{\rm k}-2g|{\bf k}|k_{0}\delta\langle\phi\rangle\\\nonumber\\
\label{56}
\delta\langle\dot\phi\rangle&=&\frac{4\pi}{(1+Z)}\delta\langle\Pi\rangle\\\nonumber\\
\label{57}
\delta\langle\dot\Pi\rangle&=&-\left(\frac{\mu^2}{4\pi}+\frac{\delta\mu^2}{4\pi}+\frac{D}{2}G(\mu)\right)\delta\langle\phi\rangle-\frac{C}{4\pi}\langle\phi\rangle^{(0)}\delta\langle\phi\rangle\nonumber\\\nonumber\\
&-&\frac{D}{8\pi}{\langle\phi\rangle^{(0)}}^2\delta\langle\phi\rangle+\frac{4g}{(2\pi)^3}\int
d^3{\rm k}^{\prime} 
\frac{|{\bf k}^{\prime}|}{{({\bf
      k}^{\prime}}^2+M^2)^{1/2}}\delta\varphi_{\rm k}\;\;. 
%
\end{eqnarray}

\noindent
Eliminating the momenta $\delta\gamma_{\rm k}$ and
$\delta\langle\Pi\rangle$ we might rewrite (\ref{54})-(\ref{57})
in a more compact second order equations:

\bea
\label{57a}
\delta\ddot\varphi_{\rm k}
&=&-4\bar\omega^{\mbox{\tiny 2}}\delta\varphi_{\rm k}
   -2g|{\bf k}|\delta\langle\phi\rangle\\\nonumber\\ 
\label{57b}
(1+Z)\delta\langle\ddot\phi\rangle
&=&-\left( \mu^2 + \Sigma \right)\delta\langle\phi\rangle\ 
   +16\pi g\int_{\bf k}h({\bf k})\delta\varphi_{\rm k} 
\eea
\noindent
where we use notations
\be
h({\bf k})=\frac{|{\bf k}|}{\bar k_0^{\mbox{\tiny}}}
\ee
with
$
\bar k_0^{\mbox{\tiny }}=\sqrt{{\bf k}^2+\bar m^2} 
$
and
\be
\label{606d}
\Sigma
\equiv \delta\mu^2+\frac{D}{2}4\pi
  G(\mu)+C\langle\phi\rangle^{(0)}
+\frac{D}{2}{\langle\phi\rangle^{(0)}}^2\;\;.  
\ee

\noindent
Thus, the small oscillation dynamics is described by coupled linear
oscillator equations as usual in RPA treatment. In particular, these
modes decouple when $g=0$ yielding two simple oscillator equations.

\indent
A solution to this problem involves determining the normal modes of
small oscillation and their frequencies. 
On the other hand, earlier studies have been demonstrated that these
elementary excitations can be interpreted as quantum particles.  
In our case here, $\delta\varphi_{\rm k}$ can be seen as two-fermion
spinless wavefunction while $\delta\langle\phi\rangle$ gives an
one-meson physics of the system in this regime. 
Furthermore, $\rm k$ stands for the relative momuntum of two fermions and
its total momentum is constant of motion for this context of uniform
system, therefore there is no explicit dynamics involved. 
Notice also that for scalar setor, the particles do not depend on the
momentum in this context. 

\renewcommand{\theequation}{3.\arabic{equation}}
\setcounter{equation}{0}
\begin{center}
{\bf III. RPA EQUATIONS AS A SCATTERING PROBLEM}
\end{center}


\indent
In the last section, we obtained the linear approximation for the gaussian
equations of motion. They describe the physics of elementary
excitations of the system. 
We shall show in this and next section that this 
coupled linear oscillator equations can be solved analytically
giving the wavefunctions and the spectrum of excitations. 

\indent 
Let us first Fourier transform the wavefunctions to the
energy representation,

\vspace{-0.5cm}
\begin{eqnarray}
\label{58}
\delta\varphi_{\rm k}(t)
&=&\int d\omega \delta\varphi_{\rm k}(\omega)e^{i\omega t} \nonumber\\ \\
\delta\langle\phi\rangle (t)
&=&\int d\omega \delta\langle\phi\rangle(\omega) e^{i\omega t} \nonumber
\end{eqnarray}

\noindent
where $\delta\varphi_{\rm k}$ and $\delta\langle\phi\rangle$ are now
energy-dependent amplitudes.  
Substituting now (\ref{58}) into (\ref{57a})-(\ref{57b}) we have
%
%
\begin{eqnarray}
\label{66a}
\left(\omega^{2}-4\bar\omega^2\right)\delta\varphi_{\rm k}(\omega)
&=&2g|{\bf  k}|\delta\langle\phi\rangle(\omega) \\\nonumber\\
\label{66b}
\left(\omega^{2}-\mu^2+Z\omega^2-\Sigma\right)\delta\langle\phi\rangle(\omega)
&=&-16\pi g\int_{\bf k}h({\rm k})\delta\varphi_{\bf k}(\omega)
%
\end{eqnarray}
\vskip 0.5cm
\noindent
Since the oscillation amplitudes play the roles of wavefunctions of
quantum particles, it is then more convenient to treat this system
as a coupled channel scattering problem with appropriate boundary
conditions \cite{KL98}. 

\indent
Next, we will discuss the scattering process where the source is a
two-fermion wave. In this case, we can solve (\ref{66b}) as follows: 

\begin{equation}
\label{66c}
\delta\langle\phi\rangle (\omega)
=\frac{-16\pi g}{\omega^2-\mu^2+Z\omega^2-\Sigma + i\eta}
      \int_{\bf k'} h({\rm k'}) \delta\varphi_{\rm k'} (\omega)\;\; .
\end{equation}

\noindent
In this expression we have included the boundary condition of there is
no incident wave of $\delta\langle\phi\rangle$.
Substituting this into (\ref{66a}) and rewriting it in terms of
a new variable
\be
\label{66d}
\Psi_{\rm k}=\sqrt{\bar k_0}\delta\varphi_{\rm k}
\ee

\noindent
one finds 
\be
\label{66e}
\left(\omega^{2}-4\bar k_0^2\right)\Psi_{\rm k}
=\frac{-32\pi g^2}{\omega^2-\mu^2+Z\omega^2-\Sigma + i\eta}
\frac{|\bf k|}{\sqrt{\bar k_0}}
\int_{\rm k'} \frac{|\bf k|}{\sqrt{\bar k_0}}\Psi_{\rm k'} \;\; ,
\ee

\noindent
where the Green's Function includes the effects of coupling of $\Psi$
to $\delta\langle\phi\rangle$. 
Notice also that the potential is separable in the sense that
\cite{Ne82} 

\begin{equation}
\label{66f}
\langle{\bf k}|V|{\bf k}^{\prime}\rangle=v({\bf k})v({\bf k'})
=\frac{|\bf k|}{\sqrt{\bar k_0}}
 \frac{|\bf k'|}{\sqrt{\bar k'_0}}
\end{equation} 

\noindent
It is now convenient to rewrite it as an integral equations,
\begin{eqnarray}
\label{67}
\!\!\!\! 
\Psi({\bf k},{\bf q};\omega)
&=&\alpha\delta({\bf q}-{\bf k})
\nonumber\\\nonumber\\ \!\!&+&\!\!
\frac{1}{[\omega^2-4\bar k^2_0 + i\epsilon ]}
\frac{-32\pi g^2}{[\omega^2-\mu^2+Z\omega^2-\Sigma +i\eta]}
v({\rm k})
\int_{\rm k'} v({\rm k'}) \Psi({\bf k}^{\prime},\bf q;\omega)\;,  
\end{eqnarray}

\noindent 
where ${\bf q}$ is the relative momentum for two incident
quasi-fermions and $\alpha$ is an overall phase factor. 
We have used the boundary condition of the outgoing wave condition
$(+i\epsilon)$ as solution of Eq.(\ref{66f}), but we could have chosen
e.g.  the incoming wave condition $(-i\epsilon)$ or Van Kampen wave
condition \cite{CTP87} or other conditions. 

\indent
The integral equation (\ref{67}) can be solved as usual. We first
multiply the expression  
In order to solve  we integrate this expression with respect
to ${\bf k}$,  

\begin{equation}
\label{68}
\int_{\rm k}v ({\bf k})\Psi({\bf k}, {\bf q};\omega)
=\frac{\alpha v({\bf q})}
      {1+\left(\frac{\displaystyle 32\pi g^2}
              {\displaystyle\omega^2-\mu^2+Z\omega^2-\Sigma+i\epsilon}\right) 
               I^{+}(\omega)}
\end{equation}
\vskip 0.5cm
\noindent
where
\be
\label{68a}
I^{+}(\omega)=\int_{\bf k}
\frac{|{\bf k}|^2}
     {\sqrt{{\bf k}^2+\bar{m}^2} (\omega^2-4\bar k_0^2 + i\epsilon )}
\ee
Substituting this back result into (\ref{67}) yields

\begin{equation}
\label{69}
\Psi({\bf k}, {\bf  q};\omega)
=\alpha\delta({\bf q}-{\bf k})
+\frac{1}{\omega^2-4\bar k_{0}^2+i\epsilon}
 \frac{\alpha v({\bf q})}{\Delta^{+}(\omega)}
\end{equation}
\vskip 0.5cm

\noindent
with

\begin{equation}
\label{70}
\Delta^{+}(\omega)
=-\frac{1}{32\pi g^2}\left( \omega^2-\mu^2+Z\omega^2-\Sigma \right)
 -I^{+}(\omega)
\end{equation}
\vskip 0.5cm
\noindent
We have, thus, found an analytical solution for the elastic channel of
two-fermion scattering problem given by (\ref{66a})-(\ref{66b}). 

\indent
The special form of interacting potential in this case allows one also
to get easily a closed expression for the scattering matrix

\begin{equation}
\label{73}
T({\bf k},{\bf k}';\omega)= v({\bf k})
\frac{1}{\Delta^{+}(\omega)}
v({\bf k}')
\end{equation}

\noindent
with $\Delta^{+}(\omega)$ given by (\ref{70}). In summary, this
section has discussed the solutions for RPA equations. These
elementary excitations describe a coupled channel scattering problem
and we have studies the particular case of two-fermion elastic
process.  
For this simple interacting potential, it is easy to obtain closed
expression for the two-fermion wavefunction and the scattering matrix,
where several dynamical behavior can be read off from
$\Delta^{+}(\omega)$.
The remained problem is the divergent integral $I^{+}(\omega)$ in
(\ref{70}) which must be removed with the help of countertems. 
We shall show in the next section that besides the counterterms used
static discussion also apply we will need a convenient choice of $Z$.

\renewcommand{\theequation}{4.\arabic{equation}}
\setcounter{equation}{0}
\begin{center}
{ \bf IV. RENORMALIZATION AND BOUND STATE SOLUTION }
\end{center}

\indent
This section will investigate the conditions for the existence of
bound states of Dirac spin-1/2 particles in a system of scalar
plasma, utilizing the framework developed in the previous sections
\cite{NTP97}. 
In this context, the standard procedure is through an analysis of the
positions of poles of the scattering matrix. The equation (\ref{70}),
however, 
has a divergent integral. We shall show next that the divergent terms
can be kept directly under control with the help of
(\ref{94})-(\ref{97}) and a 
convenient choice of $Z$, yielding a finite expression for
$\Delta^{+}(\omega)$. 

\indent 
Let us thus substitute the counterterms (\ref{94})-(\ref{97}) into
(\ref{70}), after some algebra one gets
 
\begin{equation}
\label{74}
\Delta^{+}(\omega)=\frac{1}{8\pi g^2}\left[(1+4\pi
  g^2L(m))\omega^2-\mu^2+16\pi g^2G(0)-24\pi
  g^2M^2L(m)\right]-I^{+}({\omega}) \;
\end{equation}

\noindent
with $I^{+}({\omega})$ given by (\ref{68a}). Notice that in the interval of
$0\leq\omega\leq 2M$ the integral $I_{\omega}$ is well defined and we
can take $\epsilon=0$. For $\omega>2\bar m$, on the other hand, the
spectrum is continuum. The calculation is straightforward and we find:

\be
\label{VI3}
I(\omega)= Q  - {1\over 8\pi^2}F(\omega)
-\theta(\omega-4\bar{m}^2){i\over 8\pi}\left[ \omega^2-4\bar
  m^2\right] \, ,
\ee
where 
\be
Q=\frac{1}{4\pi}\left[\Lambda^2
+\left( \frac{\omega^2}{2}-3\bar m^2 \right)\log \frac{2\Lambda}{m}
                \right]
\ee
and the finite terms are
\bea
\label{75a}
F({\omega})&=&
\left(\omega^2-6\bar m^2\right) \log\left(\frac{\bar m}{2m}\right)
+\frac{2(4\bar m^2-\omega^2)^{3/2}}{\omega}
             \tan^{-1}\sqrt{\frac{\omega^2}{4\bar m^2-\omega^2}}
\qquad 0\le\omega^2\le4\bar m^2 \nonumber\\\\
\label{75b}
F({\omega})&=&
\left(\bar m^2-6\omega^2\right) \log\left(\frac{\bar m}{2m}\right)
+\frac{2(\omega^2-4\bar m^2)^{3/2}}{\omega}
\log\frac{\omega+\sqrt{\omega^2-4\bar m^2}}{\omega-\sqrt{\omega^2-4\bar m^2}}
\qquad \omega^2> 4\bar m^2  \nonumber\\
\eea

\noindent
Comparing (\ref{74}) and (\ref{VI3}) one notes immediately that there
is still a logarithmic divergence left. This will be canceled with the
following choice of wavefunction renormalization \cite{AH88}

\be
\label{75c}
Z\equiv 4\pi g^2 L(m) \; .
\ee

\noindent
The resulting finite expression is
 
\bea
\label{76a}
\Delta^{+}(\omega)=-\frac{\pi\mu^2}{g^2m^2}+F(\omega)
-\theta(\omega-4\bar{m}^2){i\over 8\pi}\left[ \omega^2-4\bar
  m^2\right] \, .
\eea

\indent
Discussion of the problem consist now in solving the equation
\be
\label{76b}
\Delta(\omega)=0. 
\ee
Depending on the value of $\omega$ the system has different dynamical
behavior. When $\omega^2<0$ the system is unstable and for
$\omega^2>0$ we are in the scattering regime. Solution of interest
here is in the interval of $0<\omega^2<4\bar m^2$. In this case, the
system may present a stabe bound state if one finds $\omega_{_B}$ such
that $\Delta^{+}(\omega_{_B})$. The fig. 1 illustrate
$\Delta^{+}(\omega)$ for several combinations of $\mu$ and $g$ in unit
of $m$. Notice that for the same value of $g(=1)$,
$\Delta^{+}(\omega)$ has a single(none) zero when
$\mu/m<1.794(\mu/m>1.794)$. A natural way to interprete this result
is that the meson mass determines the range of the Yukawa
potential for a same coupling. When $\mu$ of the system is large, it
is more dificult to the fermions to interact and consequently
decreases the probability of forming a bound state,
and vice-versa. One can, however, compensate with inreases of values
of $g$. In order to see this, we have examinate
(\ref{76a})-(\ref{76b}) for the parameter 
space $\mu/m$ versus $g $, the result is shown in the figure 2.

\indent
In summary, this work has investigated the physics of elementary
excitations for the so called relativistic scalar plasma model. 
To reach this goal we have derived RPA equations for this sistem by
linearizing the TDHFB equations obtained in a previous work. 
In this context, the amplitudes of excitations are identified with quantum
particles and the resulting equations allow one to study scattering
processes nonperturbatively.
We have solved this RPA equation analytically using well known
procedures of scattering theory where the scattering amplitude
obtained  has a simple form.  
We have also shown that the usual definitions of countertems can be
applied to this resulting expression, from which relevant physics
about the excitations of the system can be obtained. 
In particular, our results indicate existence of bound state for
certain region of phase diagram. 

\vskip 0.5cm
\renewcommand{\theequation}{A.\arabic{equation}}
\setcounter{equation}{0}
\begin{center}
{\bf Apendix A: Mean-field kinetic equations}
\end{center}

\normalsize
\baselineskip=18pt

\indent
In this apprendix we will review briefly the results obtained of I. 
The method used there was developed earlier in the context the
nonrelativistic nuclear many-body dynamics by Nemes and de Toledo Piza
\cite{TP82}. 
More recently, our group has applied this technique to $\phi^{4}$
field theory \cite{LTP92} as well as to the Chiral Gross-Neveu model
\cite{NTP96}. 
Our approximation focus on the time evolution of a selected set of
observables, which in the case of scalar plasma system are
theexpectation values of linear, $\phi(x)$, and bilinear field
operators such as $\phi(x)\phi(x)$ , $\bar\psi(x)\psi(x)$ ,
$\psi(x)\psi(x)$ and etc. 
Because of further convenience, we work instead with expressions which
are bilinear in the creation and annihilation parts of the fields.

\indent
In the Heisenberg picture, $\phi(x)$ are scalar spin-$0$ fields 

\begin{equation}
\label{2}
\phi({\bf x},t)=\sum_{\bf p} \frac{1}{(2Vp_{0})^{1/2}}
\left[b_{\bf p}(t)e^{i{\bf p}.{\bf x}}+b_{\bf p}^{\dag}(t)
e^{-i{\bf p}.{\bf x}}\right]\;\;,
\end{equation}

\noindent
where $b_{\bf p}^{\dag}(t)$ and $b_{\bf p}(t)$ are boson creation 
and annihilation operators. For the spin-$1/2$ fields we have

\begin{eqnarray}
\label{3}
\psi({\bf x},t)&=&\sum_{{\bf 
k},s}\left(\frac{M}{k_{0}}\right)^{1/2}\frac{1}{\sqrt V}
\left[u_{1}({\bf k},s)a_{{\bf k},s}^{(1)}(t)e^{i {\bf k}.{\bf x}}
+u_{2}({\bf k},s){a_{{\bf k},s}^{(2)}}^{\dag}(t)
e^{-i {\bf k}.{\bf x}}\right]\;\;,
\nonumber\\\label{4.1} \\
\bar\psi({\bf x},t)&=&\sum_{{\bf 
k},s}\left(\frac{M}{k_{0}}\right)^{1/2}\frac{1}{\sqrt V}\left[\bar 
u_{1}({\bf k},s){a_{{\bf k},s}^{(1)}}^{\dag}(t)e^{-i {\bf k}.{\bf x}}+
\bar u_{2}({\bf k},s)a_{{\bf k},s}^{(2)}(t)
e^{i {\bf k}.{\bf x}}\right]\;\;,
\nonumber
\end{eqnarray}

\noindent
where ${a_{{\bf k},s}^{(1)}}^{\dag}(t)$ and 
$a_{{\bf k},s}^{(1)}(t)$ [${a_{{\bf k},s}^{(2)}}^{\dag}(t)$ and 
$a_{{\bf k},s}^{(2)}(t)$] are fermion creation and
annihilation operators associated with positive [negative]-energy
solutions $u_{1}({\bf k},s)$ [$u_{2}({\bf k},s)$] of Dirac's
equation. 

\indent
In order to deal with condensate and pairing dynamics of the scalar
field we define the Bogoliubov transformation as follows \cite{RS80}:

\begin{equation}
\label{4}
\left[
\begin{array}{c}
d_{\bf p}(t)\\
        \\
d_{-\bf p}^{\dag}(t)
\end{array}
\right]=\left[
\begin{array}{cc}
\cosh \kappa_{\rm p}+i\frac{\eta_{\rm p}}{2}  &  
-\sinh \kappa_{\rm p}+i\frac{\eta_{\rm p}}{2}\\ \\
-\sinh \kappa_{\rm p}-i\frac{\eta_{\rm p}}{2} &  
\cosh \kappa_{\rm p}-i\frac{\eta_{\rm p}}{2}
\end{array}
\right]
\left[
\begin{array}{c}
\beta_{\bf p}(t)  \\
      \\
\beta_{-\bf p}^{\dag}(t)
\end{array}
\right]\;\;.
\end{equation}
\vskip 0.5cm
\noindent
where $d_{\bf p}$ is the shift boson operator defined by

\begin{equation}
d_{\bf p}(t)\equiv b_{\bf p}(t)- B(t)\delta_{{\bf
p},0}\;\;\;\;{\mbox{with}}\;\;\;\; 
B_{\bf p}(t) \equiv\langle b_{\bf p}(t) \rangle = Tr_{\mbox{\tiny
BF}}\;\left[b_{\bf p}(t){\cal F}\right]\; 
\end{equation}
\vskip 0.4cm
\noindent
In the case of fermions we restrict ourselves for simplicity to Nambu
transformation, which can be parametrized in a form that incorporates
the unitarity constraints as 

\begin{equation}
\label{5}
\left[\!
\begin{array}{c}
      a_{{\bf k},s}^{(1)}    \\
                   \\
      a_{{\bf k},s}^{(2)}    \\
                   \\
{a_{-{\bf k},s}^{(1)}}^{\dag}\\
                   \\
{a_{-{\bf k},s}^{(2)}}^{\dag}
\end{array}
\!\right]\!=\!\left[\!
\begin{array}{cccc}
\cos\varphi_{\bf{k}} &      0       &      0       & 
          -e^{-i\gamma_{\bf{k}}}\sin\varphi_{\bf{k}} \\
          &              &              &              \\  
  0       & \cos\varphi_{\bf{k}} & e^{-i\gamma_{\bf{k}}}
\sin\varphi_{\bf{k}} &     0        \\ 
          &              &              &              \\ 
  0       & -e^{i\gamma_{\bf{k}}}\sin\varphi_{\bf{k}}
          &\cos\varphi_{\bf{k}} &     0        \\ 
          &              &              &              \\
e^{i\gamma_{\bf{k}}}\sin\varphi_{\bf{k}} &      0       &      0
          & \cos\varphi_{\bf{k}} 
\end{array}
\!\right]\left[
\!\begin{array}{c}
    \alpha_{{\bf k},s}^{(1)}    \\
                         \\ 
    \alpha_{{\bf k},s}^{(2)}    \\
                         \\
{\alpha_{-{\bf k},s}^{(1)}}^{\dag}\\
                         \\
{\alpha_{-{\bf k},s}^{(2)}}^{\dag}
\end{array}
\!\right]
\end{equation}
\vskip 0.5cm
\noindent

\indent
The next step is to obtain the mean-field time evolution for the mean
values of the gaussian observables in the context of the initial-value
problem.  In other words, we want the gaussian mean-field equations of
motion for the parameters $\varphi_{\bf k}(t)$ , $\gamma_{\bf
k}(t)$, $\eta_{\rm p}(t)$, $\kappa_{\rm p}(t)$, for the quasi-particle
occupation numbers $\nu_{{\rm k},\lambda}(t)$, $\nu_{\rm p}(t)$ and
for $\langle\phi\rangle$ and $\langle\Pi\rangle$. In Ref.\cite{TNa98}
we obtained for the bosonic variables 

\begin{eqnarray}
\label{8}
\dot \nu_{\rm p} &=& Tr_{\mbox{\tiny BF}}[\beta_{\bf p}^{\dag}\beta_{\bf p},H] {\cal F}\\ \nonumber\\
\label{9}
-i\dot\kappa_{\rm p}
-e^{-\kappa_{\rm p}}\left(\dot\eta_{\rm p}+\dot\kappa_{\rm p}\eta_{\rm
    p}\right) 
&=& \frac{Tr[\beta_{\bf p}^{\dag}\beta_{-\bf p}^{\dag},H] {\cal F}}
{1+2\nu_{\rm p}} \\\nonumber\\
\label{10}
i \langle\dot\phi(t)\rangle&=&Tr_{\mbox{\tiny BF}}[\phi(t),H]{\cal F}\;\;
\end{eqnarray}

\noindent
In the case of fermionic variables, we obtained
 
\begin{eqnarray}
\label{11}
\dot\nu_{{\bf k},\lambda}
&=&Tr_{\mbox{\tiny BF}}
\left([{\alpha_{{\bf k},s}^{(\lambda)}}^{\dag}\alpha_{{\bf k},s}^{(\lambda)},
H]{\cal F}\right) \\
\nonumber\\
\label{12}
\left[i\dot\varphi_{\bf k}+\dot\gamma_{\bf k}\sin\varphi_{\bf k}
\cos\varphi_{\bf k}\right]e^{-i\gamma_{\bf k}}
&=&
\frac
{Tr_{\mbox{\tiny BF}}\left([\alpha_{-{\bf k},s}^{(1)}\alpha_{{\bf k},s}^{(2)},
H]{\cal F}\right)}{1-\nu_{{\rm k},s}^{(1)}-\nu_{{\rm k},s}^{(2)}}
\;\;.
\end{eqnarray}
\vskip 0.5cm

\indent
Our implementation of the gaussian mean-field approximation consists in
approximating ${\cal F}$ by a truncated many-body density operator
${\cal F}_{0}(t)={\cal F}^{\mbox{\tiny B}}_{0}
{\cal F}^{\mbox{\tiny F}}_{0}$. 
The factorized form of the ${\cal F}_{0}(t)$ embodies what we refer 
to as the double mean field approximation. 
The subsystem densities ${\cal F}^{\mbox{\tiny B}}_{0}$
and ${\cal F}^{\mbox{\tiny F}}_{0}$ are in fact unit trace gaussian
densities, written in the form of an exponential of a  bilinear, 
Hermitian expression in the creation and annihilation parts of the
bosonic and of the fermionic fields, respectively. They can be,
however, rewritten in diagonal form, (\ref{32a}) and (\ref{32b}),
when one uses 
Bogolyubov quasiboson and quasifermion operators. With these
assumptions the equations (\ref{8})-(\ref{12}) of are closed now and
they will determinate the time rateof change of gaussian variables. 
The results are shown in (\ref{41})-(\ref{44}).

\vskip 1.0cm
\centerline{\Large\bf ACKNOWLEDGMENTS}
\vskip 0.7cm

The authors E.R.T.N. and C-Y. L. are supported by Conselho Nacional de
Desenvolvimento Cient{\'{\i}}fico e Tecnol\'ogico (CNPq), Brazil.

\vskip 1.0cm
\centerline{\large\bf Figure Captions}

\vskip 0.5cm
\noindent
Figure 1. The behavior of the function $\Delta(\omega)$ as a function
of energy in unit of $m$.

\vskip 0.5cm
\noindent
Figure 2. Existence of bound state of two fermion as a function of
parameters $\mu/m$ and $g$.

\end{document}